\documentclass[12pt,preprint]{aastex}

\def \agile {AGILE}

\def \glast {{\it Fermi}}
\def \igr {INTEGRAL}
\def \swi {{\it Swift}}

\def \ergcmsec{\hbox{erg cm$^{-2}$ s$^{-1}$}}
\def \phcmsec{\hbox{photons cm$^{-2}$ s$^{-1}$}}

\def \gray {$\gamma$-ray}
\def \source {\hbox{3C~454.3}}
\slugcomment{Accepted for publication in The Astrophysical Journal Letters}

\shorttitle{3C~454.3: the brightest gamma-ray blazar in the sky}
\shortauthors{Vercellone et al.}

\begin{document}

\title{The brightest gamma-ray flaring blazar in the sky:
AGILE and multi-wavelength observations of 3C~454.3 during November 2010}

\author{S.~Vercellone\altaffilmark{1} ,
 E.~Striani\altaffilmark{2,3},    
 V.~Vittorini\altaffilmark{4},  
  I.~Donnarumma\altaffilmark{4}, 
 L.~Pacciani\altaffilmark{4},
 G.~Pucella\altaffilmark{5},
 M.~Tavani\altaffilmark{2,3,4,6},
%
%
  C.M.~Raiteri\altaffilmark{7},
  M.~Villata\altaffilmark{7},
  P.~Romano\altaffilmark{1},  
  M.~Fiocchi\altaffilmark{4},
  A.~Bazzano\altaffilmark{4},
  V.~Bianchin\altaffilmark{8},
  C.~Ferrigno\altaffilmark{9},
  L.~Maraschi\altaffilmark{10},
  E.~Pian\altaffilmark{11,12,13},
  M.~T\"urler\altaffilmark{9},
  P.~Ubertini\altaffilmark{4},
%
%
  A.~Bulgarelli\altaffilmark{8},
  A.W.~Chen\altaffilmark{14},
  A.~Giuliani\altaffilmark{14}, 
  F.~Longo\altaffilmark{15},
%
%
  G.~Barbiellini\altaffilmark{15},
  M.~Cardillo\altaffilmark{2,4},
  P.W.~Cattaneo\altaffilmark{16},  
  E.~Del Monte\altaffilmark{4},
  Y.~Evangelista\altaffilmark{4}, 
  M.~Feroci\altaffilmark{4},
  A.~Ferrari\altaffilmark{6,17}, 
  F.~Fuschino\altaffilmark{8}, 
  F.~Gianotti\altaffilmark{8},
  M.~Giusti\altaffilmark{4}, 
  F.~Lazzarotto\altaffilmark{4},    
  A.~Pellizzoni\altaffilmark{18},
  G.~Piano\altaffilmark{4},   
  M.~Pilia\altaffilmark{18,19},   
  M.~Rapisarda\altaffilmark{5},
  A.~Rappoldi\altaffilmark{16},
  S.~Sabatini\altaffilmark{4},   
  P.~Soffitta\altaffilmark{4},
  M.~Trifoglio\altaffilmark{8},
  A.~Trois\altaffilmark{18},
%
%
  P.~Giommi\altaffilmark{20},
  F.~Lucarelli\altaffilmark{20},
  C.~Pittori\altaffilmark{20},
  P.~Santolamazza\altaffilmark{20},
  F.~Verrecchia\altaffilmark{20},
%
%
  I.~Agudo\altaffilmark{21,22},
  H.D.~Aller\altaffilmark{23},
  M.F.~Aller\altaffilmark{23},
  A.A.~Arkharov\altaffilmark{24},
  U.~Bach\altaffilmark{25},
  A.~Berdyugin\altaffilmark{26},
  G.A.~Borman\altaffilmark{27},
   R. Chigladze\altaffilmark{31},
  Yu.S.~Efimov\altaffilmark{27},
  N.V.~Efimova\altaffilmark{24,28},
  J.L.~G\'omez\altaffilmark{21},
  M.A.~Gurwell\altaffilmark{29},
  I.M.~McHardy\altaffilmark{30},
  M.~Joshi\altaffilmark{22},
  G.N.~Kimeridze\altaffilmark{31},
  T.~Krajci\altaffilmark{32},  
  O.M.~Kurtanidze\altaffilmark{31},
  S.O.~Kurtanidze\altaffilmark{31},
  V.M.~Larionov\altaffilmark{24,28,33}, 
  E.~Lindfors\altaffilmark{26},
  S.N.~Molina\altaffilmark{21},
  D.A.~Morozova\altaffilmark{28},
  S.V.Nazarov\altaffilmark{27},
  M.G.~Nikolashvili\altaffilmark{31},
  K.~Nilsson\altaffilmark{34},
  M.~Pasanen\altaffilmark{26},
  R.~Reinthal\altaffilmark{26},
  J.A.~Ros\altaffilmark{35},
  A.C.~Sadun\altaffilmark{36},
  T.~Sakamoto\altaffilmark{37},
  S.~Sallum\altaffilmark{38,39},
  S.G.~Sergeev\altaffilmark{27},
  R.D.~Schwartz\altaffilmark{40},
  L.A.~Sigua\altaffilmark{31},
  A.~Sillanp\"{a}\"{a}\altaffilmark{26},
  K.V.~Sokolovsky\altaffilmark{25},
  V.~Strelnitski\altaffilmark{38}
  L.~Takalo\altaffilmark{26},
  B.~Taylor\altaffilmark{22,41},
  G.~Walker\altaffilmark{38}
  }
\altaffiltext{1}{INAF/IASF--Palermo, Via U.~La Malfa 153, I-90146 Palermo, Italy}
    \email{stefano.vercellone@iasf-palermo.inaf.it}
\altaffiltext{2}{Dip. di Fisica, Univ. ``Tor Vergata'', Via della Ricerca Scientifica 1, I-00133 Roma, Italy}
\altaffiltext{3}{INFN--Roma ``Tor Vergata'', Via della Ricerca Scientifica 1, I-00133 Roma, Italy}
\altaffiltext{4}{INAF/IASF--Roma, Via del Fosso del Cavaliere 100, I-00133 Roma, Italy}
\altaffiltext{5}{ENEA--Frascati, Via E. Fermi 45, I-00044 Frascati (Roma), Italy}
\altaffiltext{6}{CIFS--Torino, Viale Settimio Severo 3, I-10133, Torino, Italy}
\altaffiltext{7}{INAF, Osservatorio Astronomico di Torino, Via Osservatorio 20, I-10025 Pino Torinese, Italy}
\altaffiltext{8}{INAF/IASF--Bologna, Via Gobetti 101, I-40129 Bologna, Italy}
\altaffiltext{9}{ISDC,  Universit\'e de Gen\`eve, chemin 
      d'\'Ecogia, 16 1290 Versoix Switzerland}
\altaffiltext{10}{INAF, Osservatorio Astronomico di Brera, via E. Bianchi 46, 23807, Merate, Italy}
\altaffiltext{11}{INAF, Osservatorio Astronomico di Trieste, Via G.B. Tiepolo 11, 34143 Trieste, Italy }
\altaffiltext{12}{Scuola Normale Superiore, Piazza dei Cavalieri 7, 56126 Pisa, Italy}
\altaffiltext{13}{ESO, Karl-Schwarzschild-Strasse 2, 85748 
     Garching bei M\"{u}nchen, Germany}
\altaffiltext{14}{INAF/IASF--Milano, Via E.~Bassini 15, I-20133 Milano, Italy}
\altaffiltext{15}{Dip. di Fisica and INFN, Via Valerio 2, I-34127 Trieste, Italy}
\altaffiltext{16}{INFN--Pavia, Via Bassi 6, I-27100 Pavia, Italy}
\altaffiltext{17}{Dip. di Fisica Generale, Univ. degli Studi di Torino, via P. Giuria 1, I-10125 Torino, Italy}
\altaffiltext{18}{INAF, Osservatorio Astronomico di Cagliari, localit\`{a} Poggio dei Pini, strada 54, I-09012 Capoterra, Italy}
\altaffiltext{19}{Dip. di Fisica, Univ. degli Studi dellÕ Insubria, via Valleggio 11, I-22100, Como, Italy}
\altaffiltext{20}{ASI--ASDC, Via G. Galilei, I-00044 Frascati (Roma), Italy}
\altaffiltext{21}{Instituto de Astrof\'{i}sica de Andaluc\'{i}a, CSIC, Apartado 3004, 18080, Granada, Spain}
\altaffiltext{22}{Institute for Astrophysical Research, Boston University, 725 Commonwealth Avenue, Boston, MA 02215, USA}
\altaffiltext{23}{Department of Astronomy, University of Michigan, MI, USA}
\altaffiltext{24}{Pulkovo Observatory St.-Petersburg, Russia}
\altaffiltext{25}{Max-Planck-Institut f\"{u}r Radioastronomie, Auf dem  H\"{u}gel 69, 53121 Bonn, Germany}
\altaffiltext{26}{Tuorla Observatory, University of Turku, FIN-21500 Piikki\"{o}, Finland}
\altaffiltext{27}{Crimean Astrophysical Observatory, 98049 Nauchny, Crimea, Ukraine}
\altaffiltext{28}{Astronomical Institute, St.-Petersburg State University, Russia}
\altaffiltext{29}{Harvard-Smithsonian Center for Astrophysics, MA, USA }
\altaffiltext{30}{Department of Physics and Astronomy, University of Southampton, Southampton, SO17 1BJ, United Kingdom} 
\altaffiltext{31}{Abastumani Observatory, Mt. Kanobili, 0301 Abastumani, Georgia}
\altaffiltext{32}{Astrokolkhoz Observatory, P.O. Box 1351, Cloudcroft, NM 88317, USA}
\altaffiltext{33}{Isaac Newton Institute of Chile, St.-Petersburg Branch, Russia}
\altaffiltext{34}{Finnish Centre for Astronomy with ESO (FINCA), University of Turku, FIN-21500 Piikki\"{o}, Finland}
\altaffiltext{35}{Agrupaci\'o Astron\`omica de Sabadell, Spain}
\altaffiltext{36}{Department of Physics, University of Colorado Denver, CO, USA}
\altaffiltext{37}{Center for Research and Exploration in Space Science and Technology, NASA/GSFC, Greenbelt, MD, USA} 
\altaffiltext{38}{Maria Mitchell Observatory, Nantucket, MA 02554} 
\altaffiltext{39}{Massachusetts Institute of Technology, 77 Massachusetts}
\altaffiltext{40}{Galaxy View Observatory, 102 Galaxy View Ct. Sequim, Washington 98382, USA}
\altaffiltext{41}{Lowell Observatory, Flagstaff, AZ 86001, USA}

	\begin{abstract}
Since 2005, the blazar \source{} has shown remarkable flaring activity at all frequencies, 
and during the last four years it has exhibited more than one \gray{} flare per year, becoming
the most active \gray{} blazar in the sky.
We present for the first time the multi-wavelength \agile{}, \swi{}, \igr{}, and GASP-WEBT
data collected in order  to explain the extraordinary \gray{} flare
of \source{} which occurred in November 2010.
On 2010 November 20 (MJD 55520), \source{} reached a peak flux (E$>$100\,MeV) of 
$F_{\gamma}^{\rm p} = (6.8\pm1.0)\times 10^{-5}$\,\phcmsec\, on a time scale
of about 12 hours, more than a factor of 6 higher than the flux of the brightest
steady \gray{} source, the Vela pulsar, and more than a factor of 3 brighter than
its previous super-flare on 2009 December 2--3. 
The multi-wavelength data make a thorough study of the present event possible:  the comparison with 
the previous outbursts indicates a close similarity to the one that occurred in 2009.
By comparing the broadband emission before, during, and after the \gray{} flare,
we find that the radio, optical and X-ray emission varies within
a factor 2--3, whereas the \gray{} flux by a factor of 10.
This remarkable behavior is modeled by an external Compton component
driven by a substantial local enhancement of soft seed photons.
\end{abstract}


           \section{Introduction}

The flat spectrum radio quasar \source{} (PKS 2251$+$158; z = 0.859) is the brightest 
\gray{} (0.1--10 GeV) blazar detected after the launch of the \agile{} \citep{Tavani2009:Missione}
and \glast{} \citep{Atwood2009:fermi_lat}  satellites. 
During 2007 -- 2010, AGILE 
detected and investigated several gamma-ray flares
\citep{Vercellone2010ApJ_P3, Pacciani2010ApJ_3c454, Striani2010ApJ_3c454}. 
These observations allowed us to establish a possible correlation between the \gray{} 
(0.1 --10 GeV) and the optical (R band) flux variations with no time delay, or with a
lag of the former with respect to the latter of about half a day.
Moreover, the detailed physical modeling of the spectral energy distributions (SEDs) 
when \source{} was at different flux levels provided an interpretation of the emission mechanism responsible 
for the radiation emitted in the \gray{} energy band, assumed to be inverse Compton scattering of photons from 
the broad line region (BLR) clouds off the relativistic electrons in the jet, with bulk Lorentz 
factor $\Gamma \sim 20$.
Similar results where obtained by other groups, by analyzing \glast{} and multi-wavelength data
\citep[e.g.,][]{Ghisellini2007:3C454:SED, Bonning2008:3C454_apj, Abdo2009:3C454, 
Foschini2010:3c454:variab, Bonnoli2011:3c454}.
During the period 2009 December 2--3, \source{} exhibited an intense \gray{} flare reaching a 
peak value of $F_{\gamma}^{\rm p,2009} = (2.0\pm 0.4)\times 10^{-5}$\,\phcmsec{} in the range 
0.1--10\,GeV. This extreme behavior in the \gray{} might require a 
more sophisticated modeling with respect to the widely accepted one-zone, synchrotron self-Compton (SSC)
and external Compton (EC) models \citep[][for a review of the blazar emission mechanisms and energetics]{Celotti2008:blazar:jet}. 
In particular, the almost simultaneous multi-wavelength data collected during the \gray{} flare 
required an additional population of accelerated electrons co-existing with the soft seed photons 
of the SSC/EC model \citep{Pacciani2010ApJ_3c454}.
Alternatively to the SSC/EC models, the blazar SED can be explained in the framework 
of the hadronic models \citep{Mannheim1994:hadron,Mucke2001:hadronic,Becker2008:hadron}.

Recently, \citet{Foschini2011:fsrq_variab} and \citet{Abdo2011:3c454_nov2010},
analyzing \glast{} data acquired during the November 2010 flare, show that 
the extremely fast variability, on a time-scale of about 3--6 hr,
favors a \gray{} emission originating from a compact region,
below the pc scale.

In this Letter we report on the multi-wavelength \agile{}, \swi{}, \igr{}, and GASP-WEBT
campaign covering the extraordinary November 2010 \gray{} flare of \source{}.
The quoted uncertainties are given at the
$1\sigma$ level, unless otherwise stated, and a $\Lambda$-CDM cosmology 
($h = 0.71$, $\Omega_{m}= 0.27$, and $\Omega_{\Lambda} = 0.73$) was adopted.
%

           \section{Data Analysis}\label{sect:data}

%
%
\agile{} detected increased and prolonged \gray{} emission from \source{} starting
from 2010 October 28 (MJD 55497), with a maximum emission on 2010 November  20
\citep[MJD 55520,][]{Vercellone2010:3c454ATelnov2010,
Striani2010:3c454:ATel3034,Striani2010:3c454:ATel3043,Striani2010:3c454:ATel3049}.
Level--1 AGILE-GRID data were analyzed using the AGILE Standard Analysis
Pipeline \citep[see][for a description of
the AGILE data reduction]{Vercellone2010ApJ_P3}. We used \gray{} events 
filtered with the \texttt{FM3.119} \agile{} Filter pipeline.
Counts, exposure, and Galactic background \gray{} maps were created with
a bin-size of $0.\!\!^{\circ}5 \times 0.\!\!^{\circ}5$\,,
$E \ge 100$\,MeV, and including all events collected up to $60^{\circ}$ off-axis.
To reduce the \gray{} Earth albedo contamination, we rejected all \gray{} events 
whose reconstructed directions form angles with the satellite-Earth vector $<85^{\circ}$.
We used the latest version (\texttt{BUILD-20}) of the Calibration files
(\texttt{I0010}), which will be publicly available at the ASI Science Data Centre
(ASDC) site\footnote{\url{http://agile.asdc.asi.it}},
and of the \gray{} diffuse emission model \citep{Giuliani2004:diff_model}.
We ran the AGILE Multi-Source Maximum Likelihood Analysis (\texttt{ALIKE}) 
task with an analysis radius of 10$^{\circ}$.

\begin{figure}
\epsscale{0.90}
\plotone{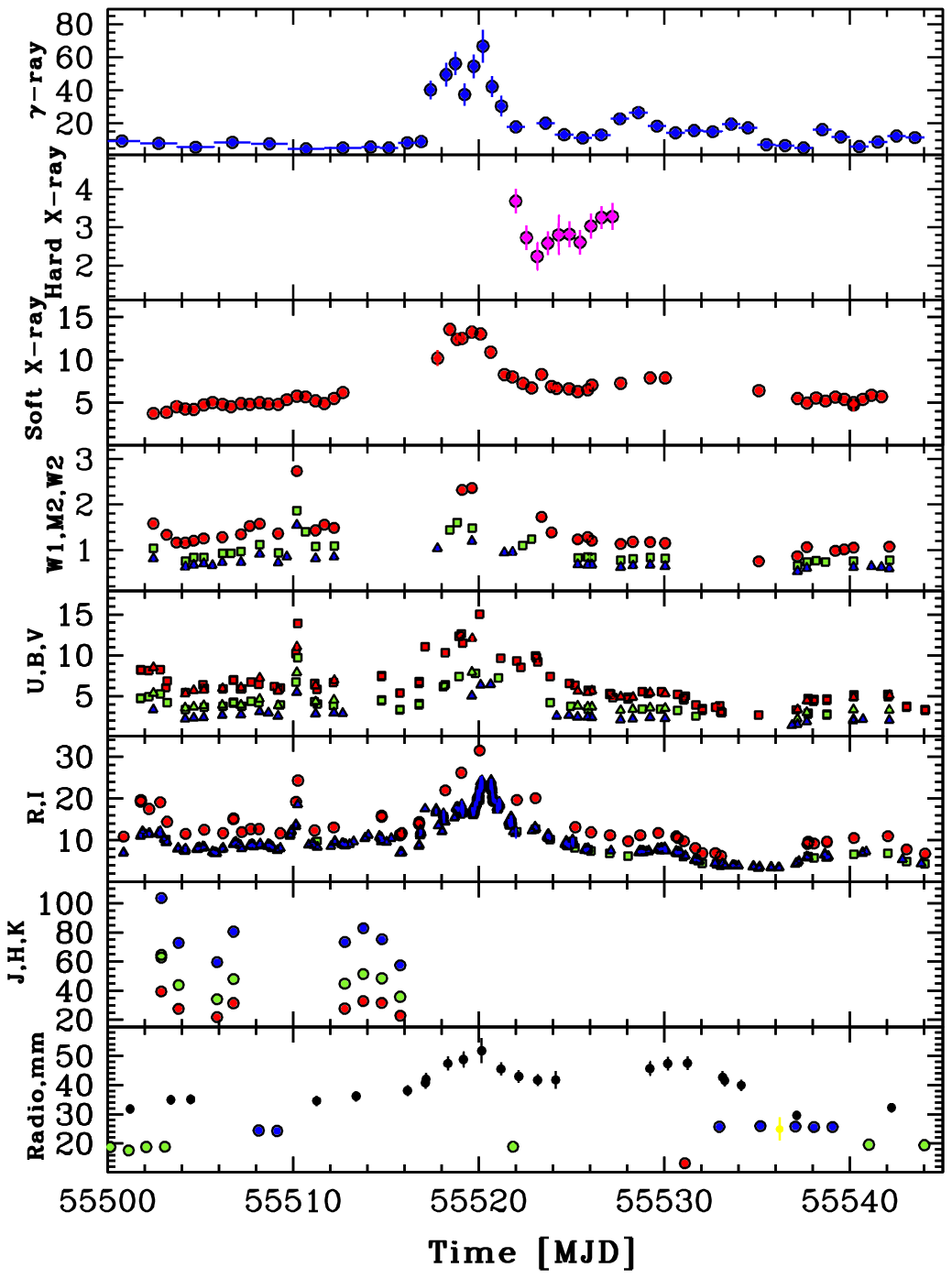}
\caption{Multi-wavelength light curves from radio to \gray. \gray{} (E$>$100\,MeV) data are in units of
$10^{-6}\,$\phcmsec; hard X-ray (20--100\,keV) data are in units of counts s$^{-1}$; soft X-ray (2--10\,keV) data are in units 
of $10^{-11}$\,\ergcmsec; UV, optical and NIR data are in units of mJy; radio data are in units of Jy. See \S\,\ref{sect:results} for details. The starting date, MJD~55500, corresponds to 2010-10-31 00:00 UT.\label{fig1}}
\end{figure}

%
%
The \swi{} X-ray Telescope (XRT) data were processed with standard procedures
({\sc xrtpipeline} v0.12.6 within {\sc HEASOFT} V.6.10), filtering and screening criteria.
The XRT data were in windowed timing mode (WT). Spectra were extracted
on an orbit-by-orbit basis due to a rotation of the roll angle, within
circular regions with a radius of 47.15 arcsec.
We used the latest spectral redistribution matrices (20101206).
The \swi{} Ultraviolet and Optical Telescope (UVOT) data analysis was performed 
using the {\tt uvotimsum} and {\tt uvotsource} tasks.
Source counts were extracted from a circular region with a 5 arcsec radius.
The background was extracted from a nearby source-free circular regions. The reported fluxes 
are on the UVOT photometric system described in \citet{Poole2008:UVOT}, and are not corrected
for Galactic extinction.

%
%
The \igr{} data were obtained as a Target of Opportunity (ToO) observation triggered immediately after
the \gray{} flare \citep[for preliminary results see][]{Pian2010:ATel_3c454_intergral}, and 
were processed using the ISDC Off-line Scientific Analysis software
\citep[\texttt{OSA V.\/9.0},][]{Courvoisier:2003:isdc}.
\igr{} observed the source between 2010 November 21.70 UT and November
27.24 UT, for a total on source time of 400 ks.
Light curves and spectra were extracted for each individual science window (SCW) and later combined. 
We investigated the 4 -- 26~keV (JEM-X both units) and 20 -- 200~keV (IBIS/ISGRI) energy ranges.
We extracted the light-curve from IBIS/ISGRI in the 20--40\,keV, and 40--100\,keV energy bands
to verify possible variability on time scales from hours to one day: the average count rate in the softer 
(harder) band
is $1.6\pm0.1$\,counts s$^{-1}$ ($1.4\pm0.1$\,counts s$^{-1}$), or about 11 (15)\,mCrab 
without significant variability. 
 We then analyzed the averaged spectrum using 8 pre-defined energy channels in JEM-X
and 16 customized energy channels for IBIS/ISGRI with {\sc Xspec}
(v. 12.6.0). We assumed a single power law model 
and free cross-calibration constants for JEM-X ($C_{1}$ and $C_{2}$) with respect to ISGRI,
and a 3\% systematic error to account for the instrumental calibration uncertainties. 
We obtain (uncertainties at 90\% c.l.):
$\Gamma=1.59\pm0.08$\, ($\chi^2_\mathrm{red}/\mathrm{d.o.f.}=1.35/23$), $C_1=1.1\pm0.2$\,, $C_2=1.3\pm0.2$\,,
$F_\mathrm{4-200\,keV}=\left(4.7^{+0.2}_{-0.4}\right)\times10^{-10}$\,\ergcmsec.

\begin{figure}
\epsscale{0.90}
\plotone{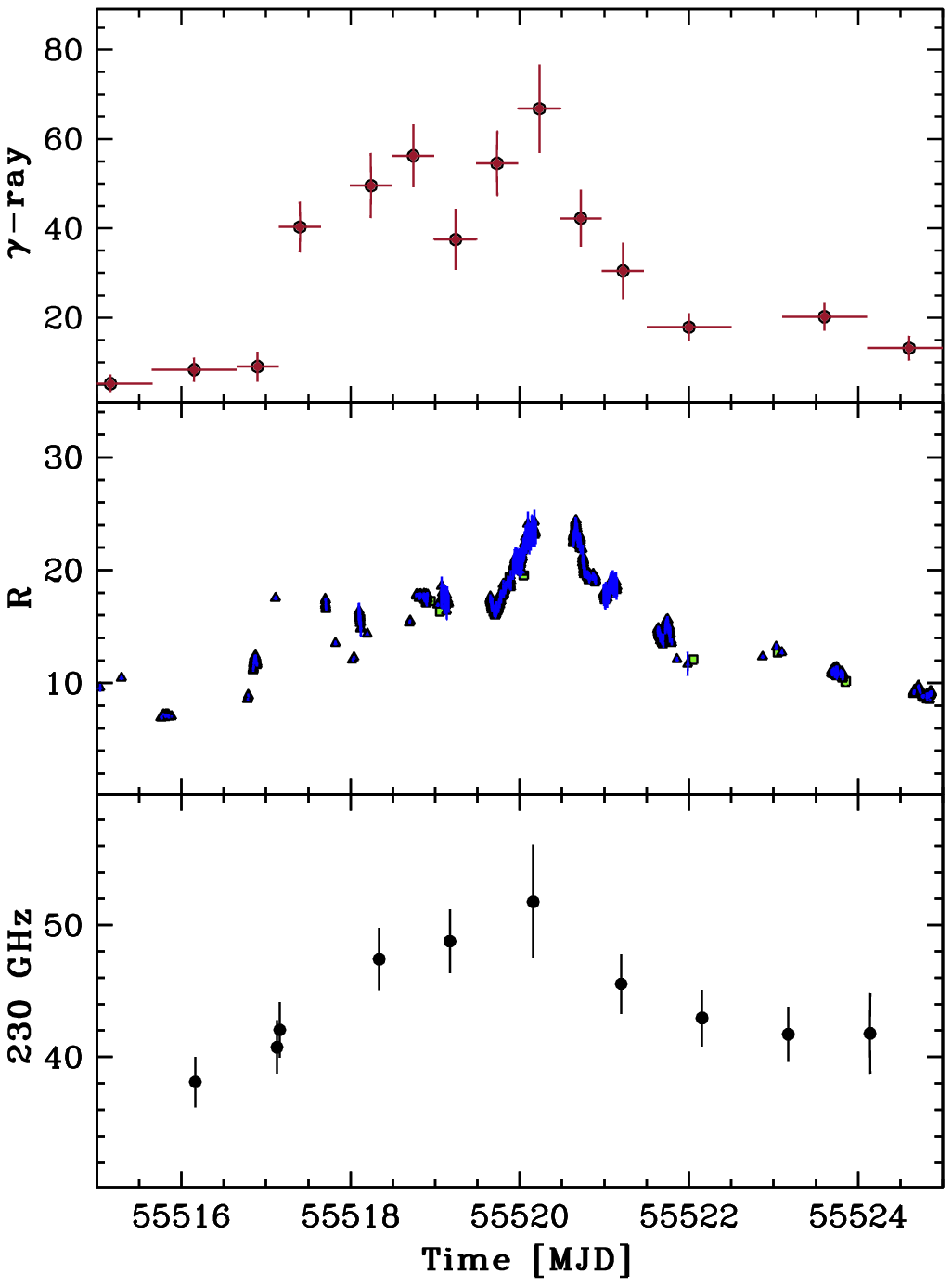}
\caption{Zoom of the \gray{} (top panel), $R$-band (middle panel), and 230~GHz
(bottom panel) light curves. Units and energy bands are the same as in Figure~\ref{fig1}.
The starting date, MJD~55515, corresponds to 2010-11-15 00:00 UT.\label{fig2}}
\end{figure}

%
%
The GLAST-AGILE Support Program (GASP; \citealt{vil08,vil09}) is a project born from the Whole 
Earth Blazar Telescope\footnote{\url{http://www.oato.inaf.it/blazars/webt/}} (WEBT) in 2007,
which has analysed the multi-frequency behavior of \source{} since the
unprecedented optical outburst of 2005 \citep{vil06,vil07,rai07,rai08a,rai08b,vil09,rai11}.
The $R$-band GASP observations of \source{} in the period considered in this paper were performed 
by the following observatories:
Abastumani, Calar Alto, Crimean, Galaxy View, Goddard (GRT), Lowell (Perkins),
Maria Mitchell, New Mexico Skies, ROVOR, Roque de los Muchachos (KVA and Liverpool), Sabadell,
and St. Petersburg.
Additional data were taken by the AAVSO\footnote{\url{http://www.aavso.org/}} \citep{Krajci2010:ATel_3c454}
and the Yale Fermi/SMARTS project\footnote{\url{http://www.astro.yale.edu/smarts/glast/}}
\citep{Chatterjee2011:SMARTS}.
Some of the above observatories also provided data in the $B$, $V$, and $I$ bands.
Near-IR (NIR) data in $J$, $H$, and $K$ bands are from Campo Imperatore.
Radio flux densities were measured at:
Submillimeter Array \citep[SMA, 345 and 230 GHz, see][]{gur07},
Medicina (8 GHz), and UMRAO (14.5, 8.0, and 4.8 GHz).
Data reduction and analysis follow \cite{rai08a}.

           \section{Results}\label{sect:results}

Figure~\ref{fig1} shows the multi-wavelength light curves from radio to \gray; 
from top to bottom: \agile{}/GRID (E$>$100~MeV) at a variable time-bin of 
$\approx 2, 1$, and $0.5$--day; \igr{} (20--100~keV);
\swi{}/XRT (2--10~keV); \swi{}/UVOT UV $w1$ (red circles), $m2$ (green squares), 
and $w2$ (blue triangles); \swi{}/UVOT (triangles), and GASP-WEBT (squares)
optical $U$ (blue points), $B$ (green points), and $V$ (red points); 
GASP-WEBT optical $R$ (blue triangles, with additional data from 
AAVSO as green squares) and $I$ (red circles);
GASP-WEBT near infra-red $J$ (red points), $H$ (green points), and $K$ (blues points); 
and GASP-WEBT 4.8~GHz (red circles), 8~GHz (green circles), 
14.5~GHz (blue circles), 230~GHz (black circles), and 345~GHz (yellow circles).

When data are available, a clear peak is present at approximately MJD~55520 in all light curves.
In particular, the \gray{} light curve shows different behavior  during the pre- and the post-flare
periods, the former being much steadier than the latter, on almost the same time-scale.
The UV and optical light curves show a very similar trend, in particular the remarkably
fast flare centered at approximately MJD~55510, with rising and falling of about a factor of 2--2.5 in 
about 48 hours. During the same period, the X-ray flux varied by about 20\%, while no
significant variability is detected at other wavelengths. We also note that there is an
average offset between the GASP-WEBT and \swi{}/UVOT data in the $\it V$ and $\it B$ band.
If ($V$, $B$) and ($v$, $b$) are the GASP-WEBT and \swi{}/UVOT data, respectively,
then a good agreement is found when $B-b=0.1$\,mag, and $V-v=-0.05$\,mag. \cite{rai11}
discuss the nature of this offset and provide a tool for the calibration of the \swi{}/UVOT data. 
Figure~\ref{fig2} shows a zoom, centered on the \gray{} flare date, of the
\gray{}, $R$-band and 230~GHz light curves. We note that,
to the sudden enhancement (about a factor of 4) of the gamma-ray
flux on MJD~55517, does not correspond a similar flare in the other wavebands.

\begin{figure}
\epsscale{0.90}
\plotone{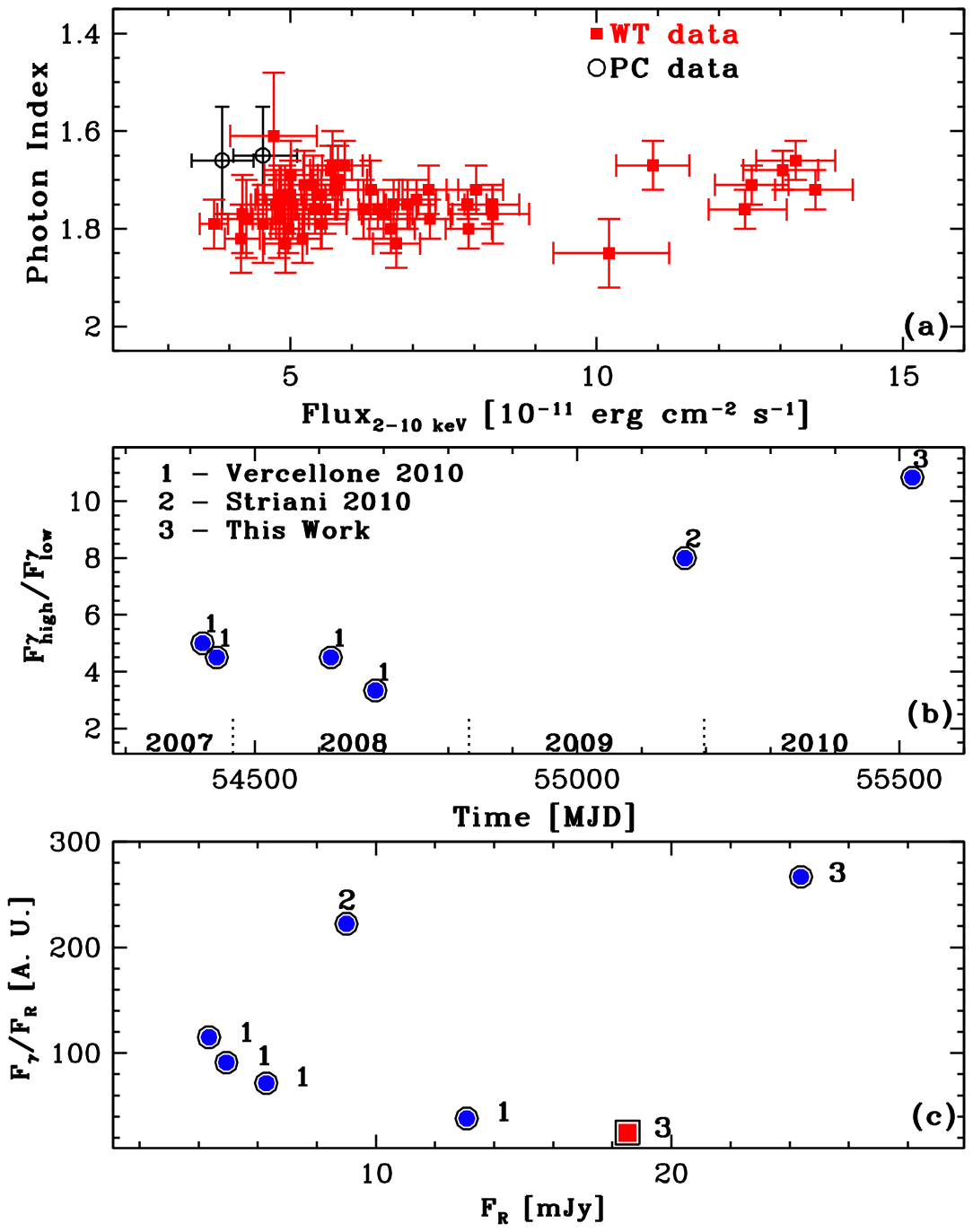}
\caption{
{\it Panel (a): }X-ray photon index as a function of the X-ray flux in the energy band 2--10 keV. 
{\it Panel (b): }peak over low \gray{} flux ratio for the major \agile{} detected flares as a function of time.
{\it Panel (c): }blue circles represent the ratio between the \gray{} and $R$-band peak fluxes as a function 
of the $R$-band peak flux,  while red square represents the same quantity for the fast optical--UV flare 
on MJD 55510.\label{fig3}}
\end{figure}

The \swi{}/XRT spectra were rebinned to have at least 20 counts per energy bin,
and were fit with an absorbed power law model.
Following \cite{Vercellone2010ApJ_P3}, the Galactic absorption was fixed to the value of
$N_{\rm H}^{\rm Gal} = 1.34 \times 10^{21}$\,cm$^{-2}$ \citep{vil06}.
In Figure~\ref{fig3} (a),  we show the X-ray photon index as a function of the
2--10\,keV X-ray flux. No particular trend is present, contrary
to what was previously reported in \cite{Vercellone2010ApJ_P3}
for the period 2007--2009.

The \gray{} spectra were computed in four different time periods, pre-flare
(MJD 55497.60 -- 55516.40), flare (MJD 55518.25 -- 55520.25),
post-flare (MJD 55521.25 -- 55543.75) and whole period. 
Table~\ref{tbl1} shows the spectral parameters that we obtained by
fitting the \agile{} data with a simple power-law:
\begin{equation}\label{eqn:def:spectra}
  F(E) =
  k \times
  \left( \frac{{\rm E}}{1\, {\rm MeV}}\right)^{-\alpha}\,.
\end{equation}
Table~\ref{tbl1} reports the numerical values during the different time periods.
We note that the flux during the flare is about a factor of 25 larger than the average value reported 
in the First \agile{} Catalogue \citep[][July 2007 - June 2008]
{Pittori2009:AGILE_Cat1}.
 %

           \section{Discussion}\label{sect:discuss}

Since its launch in 2007, \agile{} detected significant 
\gray{} emission from \source{} with repeated and prolonged flaring activity. 
%
Figure~\ref{fig3} (b),  shows the ratio 
$F^{\gamma}_{\rm high}/F^{\gamma}_{\rm low}$ as a function of time, where
$F^{\gamma}_{\rm high}$ is the flux maximum value during each flare detected
by \agile{}, and $F^{\gamma}_{\rm low}$ is the
lowest flux point just before the corresponding rising in the light curve.
While in 2007--2008 this ratio is of the order of 3--5, 
in 2009 and 2010 this ratio is at least a factor of 2--3 larger.
It is worth noting that, during 
this latest flare, the 230~GHz, optical-UV, X-ray, and \gray{} flux variations are 
almost simultaneous. 
The 230~GHz flux shows a more prolonged active state after
the \gray{} super-flare ($\sim 10$ days), while the UV and optical fluxes reach 
levels comparable or lower than the pre-flare ones.
The extremely large \gray{} dynamic range in the 2010 flare, the high 230~GHz flux
level, and the relatively bright state in the optical band, when compared with other
\gray{} flares, could not be explained only in terms of the alignment of different regions
of the jet as suggested in \cite{Vercellone2010ApJ_P3}.
Figure~\ref{fig3} (c), provides further evidence of the different behavior 
of the 2009 and 2010 flares with respect to the previous ones. 
The plot of the ratio between the \gray{} and $R$-band peak fluxes as a function of the $R$-band 
peak flux (blue circles) shows a difference among flares in the different epochs. 
The red square represents the point relative to the fast optical--UV flare on MJD 55510.
This last point is particularly interesting because of the fast rising and decay time scales ($\sim$1\,d).
To explain the lack of a simultaneous \gray{} flare,
a magnetic field enhancement by a factor of 2, a local $\gamma$-$\gamma$ absorption,
or a locally seed-photon starved zone are required.
%

%
%
\begin{deluxetable}{lrrrrr} 	
  \tablecolumns{6}
  \tabletypesize{\normalsize}
  \tablecaption{{\it Top: }\gray{} spectral parameters. {\it Bottom: } parameters for the pre-flare and flare, 
  SED models, respectively (see~\S\,\ref{sect:discuss} for details).\label{tbl1}} 	
  \tablewidth{0pt}
  \tablehead{
    \colhead{Parameter} & \colhead{Pre-flare} & \colhead{Flare} & \colhead{Post-flare} & \colhead{Whole period} 
    & \colhead{Units}}
  \startdata
\cutinhead{\gray{} spectral parameters}
  F$_{\rm E>100\,MeV}$  &  $6.2\pm0.6$        &  $53.5\pm3.7$       &  $13.8\pm0.6$      &  $16.8\pm0.7$      &  $10^{-6}$\,ph\,cm$^{-2}$s$^{-1}$\\
  $k$                                    &  $0.4$                    &  $11.1$                   &  $10.0$                  &  $3.8$                    &  $10^{-3}$\,ph\,cm$^{-2}$s$^{-1}$MeV$^{-1}$\\
  $\alpha$                           &  $1.93 \pm 0.20$  &  $2.13 \pm 0.13$  &  $2.37 \pm 0.08$  &  $2.15 \pm 0.08$  &  \\
\cutinhead{SEDs model parameters}
  $\alpha_{\rm l}$              & $2.35$                    & $2.35$                   &                     &                                & \\
  $\alpha_{\rm h}$             & $4.2$                      & $4.8$                     &                     &                               & \\
  $\gamma_{\rm min}$      & $50$                      & $80$                      &                      &                               & \\
  $\gamma_{\rm b}$          & $650$                    & $700$                    &                     &                                & \\
  $K$                                   & $300$                    & $700$                    &                     &                                & cm$^{-3}$ \\
  $R_{\it blob}$                  & $7.0$                     & $3.6$                     &                      &                               & 10$^{16}$\,cm\\
  $B$                                   & $0.65$                   & $1.1$                     &                      &                                & G\\
  $\delta$                            & $34.5$                   & $34.5$                   &                      &                                & \\
  $L_{\rm d}$                      & $2$                        & $2$                         &                      &                                & 10$^{46}$\,erg\,s$^{-1}$\\
  $T_{\rm d}$                      & $10^{4}$               & $10^{4}$               &                       &                                & $^{\circ}K$\\
  $r_{\rm d}$                       & $0.05$                   & $0.05$                  &                       &                                & pc\\
  $\Theta_{0}$                   & $1.15$                   & $1.15$                   &                       &                                & degrees\\
  $\Gamma$                      & $20$                      & $20$                      &                       &                                & \\
  \enddata
\end{deluxetable} 

The \igr{} 20--100\,keV light curve shows another
peculiarity of the November 2010 flare with respect to the intense optical and X-ray flare
of May 2005. During the November 2010 flare, the source shows only moderate 
flux variability ($\approx 1.6\sigma$ level) and a relatively hard spectrum ($\Gamma = 1.59\pm0.08$),
while in 2005, during a comparably long observation at a comparable flux level,
the source exhibited remarkable flux variability and a softer spectrum 
\citep[$\Gamma = 1.8\pm0.1$,][]{Pian2006:3C454_Integral}.
Despite the moderate variability in the 20--100\,keV band, Figure~\ref{fig1} seems to show a 
similar trend between the light curves in the soft, hard X-ray and \gray{} bands.

\begin{figure}
\epsscale{0.90}
\plotone{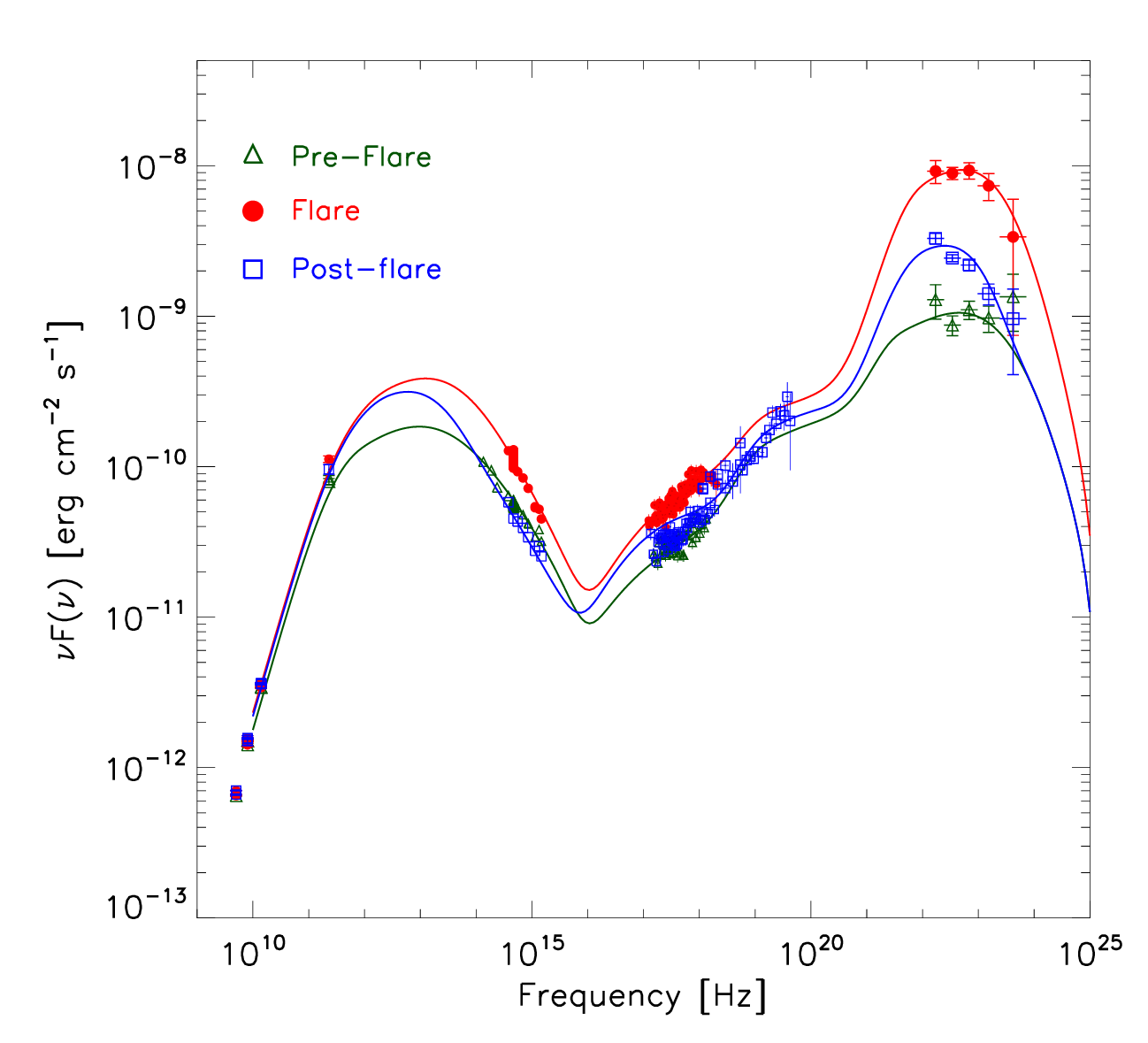}
\caption{SEDs constructed before (green open triangles), during (red filled circles), 
and after (blue open squares) the flare. The NIR, optical, UV, and X-ray data 
are corrected for Galactic extinction. The fit lines and model parameters are described 
in \S\,\ref{sect:discuss} and in Table~\ref{tbl1}. \label{fig4}}
\end{figure}

Figure~\ref{fig4} shows SEDs chosen at three epochs 
with the largest number of data points.
The NIR, optical, UV, and X-ray data are corrected for Galactic extinction. 
Green open triangles represent the pre-flare
SED, accumulated on MJD 55512. In order to obtain enough statistics, the \agile{} data were acquired 
in the period MJD 55497.60 -- 55516.40. Red filled circles represent the flare SED, accumulated on
MJD 55519 (\agile{} data:  MJD 55518.25 -- 55520.25). 
Blue open squares represent the post-flare SED, accumulated on
MJD 55526 (\agile{} data:  55521.25 -- 55543.75).  
The \igr{} IBIS/ISGRI integrated spectrum (MJD 55521.70 -- 55527.30) is also reported.

We fit the pre--flare SED by means of a one-zone leptonic model, 
considering the contributions from SSC and from external seed photons originating both from the
accretion disk (AD) and from the BLR (detailed description of this model
is given in \citealt{Vittorini2009:ApJ:Model}).
Indeed, emission from both the BLR and the AD were detected during faint states
of the source \citep{rai07}.
The solid lines represent the total contributions before (green), during (red), and after
the flare (blue).

Hadronic models are challenged by the detection of correlated variability at different 
wavelength, as pointed out by \cite{Finke2010:3C454}, and as observed by both 
\glast{} and \agile{} \citep[but see][]{Barkov2010:flare_redSG}.

The emission along the jet is assumed to be produced in
a spherical blob with comoving radius $R_{\rm blob}$ by
accelerated electrons characterized by a comoving broken power
law energy density distribution,
\begin{equation}
n_{e}(\gamma)=\frac{K\gamma_{\rm b}^{-1}} 
{(\gamma/\gamma_{\rm b})^{\alpha_{\rm l}}+
(\gamma /\gamma_{\rm b})^{{\alpha}_{\rm h}}}\,,
\label{eq:ne_gamma}
\end{equation}
where $\gamma$ is the electron Lorentz factor varying
between $80<\gamma<8\times10^{3}$, $\alpha_{\rm l}$ and
$\alpha_{\rm h}$ are the pre-- and post--break electron distribution
spectral indices, respectively, and $\gamma_{\rm b}$ is the
break energy Lorentz factor. We assume that the blob
contains an homogeneous and random magnetic field
$B$ and that it
moves with a bulk Lorentz Factor
$\Gamma$ at an angle
$\Theta_{0}$  with respect to the line of sight. The
relativistic Doppler factor is $\delta = [ \Gamma \,(1 - \beta
\, \cos{\Theta_{0}})]^{-1}$, where $\beta$ is the blob bulk
speed in units of the speed of light.
Our fit parameters are listed in Table~\ref{tbl1}.

Our modeling of the \source{} high-energy emission is based
on an inverse Compton  (IC) model with two main sources of external
seed photons:
{(1)} the AD characterized by a blackbody spectrum 
peaking in the UV with a bolometric luminosity  $L_{\rm d}$
for an IC-scattering blob at a distance $r_{\rm d}$ 
from the central part of the disk\footnote{We find 
$r_{\rm d} = 0.05$\,pc, to be compared with $r_{\rm s}=1.5\times10^{14}$\, cm
\citep[assuming $M_{\rm BH}=5\times 10^{8}$\,M$_{\odot}$,][]{Bonnoli2011:3c454}. 
Our distance value may challenge models in which
the dissipation region distance is $>$\, a few pc \citep[e.g., ][]{Jorstad2010:3C454MW}.};
{(2)} the BLR with a spectrum peaking in the $V$,
placed at a distance from the central black hole of $r_{\rm BLR}=3 \times 10^{18}$\,cm, 
and assumed to reprocess $10\%$ of the irradiating
continuum (see \citealt{Vercellone2008:3c454:ApJ_P1} for the first application of this model to 
\source{}, and \citealt{Pacciani2010ApJ_3c454} for the December 2009 flare modeling).

The flaring behavior in the optical and \gray{} energy bands is puzzling. 
We observe a first rapid flare at MJD~55510 with the optical flux rising and 
falling by a factor 2 in about 24 hours without a \gray{} counterpart. 
After a time T=7\,d, at MJD~55517, the optical flux increases by a factor of 2, followed after 
about 12 hours  by a \gray{} increase by a factor of 4.
On the contrary, on MJD~55520, the flux variations in the
optical and \gray{} bands are of the same magnitude, as expected from an EC mechanism. 
This complex behavior challenges the idea of a uniform external photon field.
A possible explanation is that an energetic particle ignition,
taking place in an average-density photon region,
causes the first optical flare at MJD~55510. Subsequently, the blob moves away by 
$c\,T\delta /(1+z)\approx 3.4\times10^{17}$\,cm towards a region with a denser external photon 
field in which a doubling in the optical flux can be followed by a stronger EC counterpart, 
as observed during the \gray{} enhanced emission at MJD~55517.
Thereon, since the blob is moving in a region with enhanced density of external 
seed photons, the optical and \gray{} flux variations have similar dynamic range
(as observed at MJD~55520), until the blob leaves this denser region. 
Alternatively, to explain the whole behavior, we can invoke a modest variation of $\Gamma$.
Subsequently, as observed in the post-flare SED, 
the \gray{} emission decreases because of 1) the radiative cooling, and 2) the decrease
of the external photon field due to the blob escaping the enhanced density region. 
The post-flare SED parameters, therefore, are similar to the flare ones, once evolution by radiative cooling
is taken into account.
During the flare, the total power carried in the jet, $P_{\rm jet}$, defined as
\begin{equation}
P_{\rm jet} = L_{\rm B} + L_{\rm p} + L_{\rm e} + L_{\rm rad}\,\,\,{\rm
  erg}\,{\rm s}^{-1},
\label{eq:Pjet}
\end{equation}
\citep[see also][]{Ghisellini2001:JetPower} where $L_{\rm B}$, $L_{\rm p}$, $L_{\rm e}$, and $L_{\rm rad}$ 
are the power carried by the magnetic field, the cold protons\footnote{If considering also the contribution of
relativistic protons, with $\langle \gamma \rangle = 10^{2}$\,, we obtain 
$P_{\rm jet} \approx 10^{49}$\,erg\,s$^{-1}$.} , the relativistic electrons, and the produced radiation, 
respectively, is  of the order of  $P_{\rm jet} \approx 10^{47}$\,erg\,s$^{-1}$.
%

We can now discuss the absence of a harder-when-brighter trend in the 2--10~keV 
energy band. During the 18-months \agile{} campaign, 
\cite{Vercellone2010ApJ_P3} found a clear trend, in particular for fluxes 
above (1--2)$\times 10^{-11}$\,\ergcmsec\,.  
\cite{Donnarumma2011:intergral_PoS} show that a behavior similar to the November 2010 one
was already present during the 2009 December 2--3 \gray{} flare
\citep[see also][]{Pacciani2010ApJ_3c454}.
We can describe the harder-when-brighter trend in terms of a dominant contribution
of the EC off the disk seed photons, EC(Disk), over the SSC component, probably
due to an increase of the accretion rate. We note that an increase
of $n_{\rm e}$ and/or $\gamma_{\rm b}$ would cause a softer-when-brighter trend, inconsistent with
our findings.
The constant X-ray photon index during the extreme \gray{} flares in 2009 and 2010 can be 
interpreted in terms of a balance of the SSC contribution with respect to the EC(Disk).
If we assume that $\gamma_{\rm b}$ increases significantly with respect to the 2007--2008
ones ($\gamma_{\rm b} = 200-300$ in 2007--2008, $\gamma_{\rm b} = 700-800$  in 2009--2010),
we obtain both an increase of the EC(Disk) component (and the shift of the peak of its
emission to higher frequencies), and a simultaneous increase of the SSC. The net result
is a roughly achromatic increase of the X-ray emission.
%

           \acknowledgments

We thank the Referee for useful comments. We thank A.~P. Marscher and S.~G. Jorstad
for Perkins and Liverpool Telescopes optical data.
We acknowledge financial contribution from: agreement ASI-INAF {I/009/10/0},
ASI contract {I/089/06/2}, RFBR Foundation grant 09-02-00092, MICIIN grant
AYA2010-14844, CEIC grant P09-FQM-4784, NSF grant AST-0907893,
NASA Fermi GI grants NNX08AV65G and NNX10AU15G, NSF/REU grant AST-0851892,
the Nantucket Maria Mitchell Association, and grant GNSF/ST08/4-404.

{\it Facilities:} \facility{AGILE}, \facility{INTEGRAL}, \facility{{\it Swift}}, 
\facility{WEBT}, \facility{AAVSO}, \facility{SMA}.

\end{document}